\documentclass[useAMS,usenatbib,a4paper]{mn2e}

\usepackage{aas_macros, multirow, mathrsfs}
\usepackage[fleqn]{amsmath}
\usepackage{tabularx}
\usepackage{graphicx}

\newcommand{\nh}{$N_{\mathrm{H}}$}
\newcommand{\erg}{ \rm \, erg \, s^{-1} cm^{-2}}

\def\calc{{\cal C}}
\def\gs{\,\rm g\,s^{-1}}

\def\msun{\ensuremath{M_{\odot}}}
\def\mdot{\ensuremath{\dot{M}}}

\def\mext{\ensuremath{\mdot_{\rm ext}}}
\def\mcrit{\ensuremath{\dot{M}_{\rm crit}}}
\def\medd{\ensuremath{\mdot_{\rm Edd}}}
\def\ledd{\ensuremath{L_{\rm Edd}}}

\def\rxte{\textsl{RXTE}}

\title[Revisiting a fundamental test of the DIM]{Revisiting a fundamental test of the disc instability model for X-ray binaries}
\author[M. Coriat et al.]{M. Coriat,$^{1}$\thanks{E-mail:
m.coriat@soton.ac.uk} R. P. Fender$^{1}$ and G. Dubus$^{2}$\\
$^{1}$School of Physics and Astronomy, University of Southampton, Highfield, Southampton, SO17 1BJ, UK\\
$^{2}$UJF-Grenoble 1/CNRS-INSU, Institut de Plan\'etologie et d'Astrophysique de Grenoble (IPAG) UMR 5274, 38041 Grenoble, France\\}

\begin{document}

\date{\today}

\pagerange{\pageref{firstpage}--\pageref{lastpage}} \pubyear{2012}

\maketitle

\label{firstpage}

\begin{abstract}

\noindent We revisit a core prediction of the disc instability model (DIM) applied to X-ray binaries. The model predicts the existence of a critical mass transfer rate, which
depends on disc size, separating transient and persistent systems. We therefore selected a sample of 52 persistent and transient neutron star and black hole X-ray binaries and verified
 if observed persistent (transient) systems do lie in the appropriate stable (unstable) region of parameter space predicted by the model.
 We find that, despite the significant uncertainties inherent to these kinds of studies, the data are in very good agreement with the theoretical expectations. We then discuss
 some individual cases that do not clearly fit into this main conclusion.
 Finally, we introduce the transientness parameter as a measure of the activity of a source and show a clear trend of the average outburst recurrence time to decrease with transientness in agreement
 with the DIM predictions. 
We therefore conclude that, despite difficulties in reproducing the complex details of the lightcurves, the DIM succeeds to explain the global behaviour 
of X-ray binaries averaged over a long enough period of time.

\end{abstract}


\section{Introduction}

Low mass X-ray binaries (LMXB) are bright X-ray sources consisting of an accreting black hole/neutron star primary and a Roche lobe filling, low-mass secondary star. Matter is transferred from the secondary star via the inner-Lagrange (L1) point and forms an accretion disc around the primary. In some cases the accretion disc undergoes sporadic outbursts which are thought to be triggered by a thermal-viscous instability, resulting in an increased mass accretion rate onto the primary and a rapid X- ray brightening.
This instability  is thought to arise when the disc effective temperature is close to $10^4$ K, enough for hydrogen to become partially ionized and for opacities to depend strongly on temperature \citep{meyer81}. The model describing the global properties of this instability applied to accretion discs is referred to as the Disc Instability Model  \citep[DIM; See][for a review of the model]{lasota01}. 

The DIM has been shown to reproduce the broad observational properties of black holes and neutron stars outbursts when it is modified to include irradiation heating and a transition to a radiatively inefficient accretion flow below a given mass transfer rate \citep{king98, dubus01}. However, several features of outburst lightcurves remain unexplained (glitches, plateaux, rebrightenings, etc.) and the model rests on critical assumptions for angular momentum transport (alpha viscosity) and unknown physics (efficient/inefficient disc transition). Hence, it is difficult to use outburst lightcurves to provide definitive tests of the DIM. In contrast, one of the core predictions of the DIM, which depends little on the details of angular momentum transport, is the existence of a critical mass transfer rate that depends on disc size, separating stable and unstable discs. A basic test is therefore to verify if observed persistent (transient) systems do lie in the appropriate stable (unstable) region of parameter space predicted by the DIM - modulo the effects of irradiation.

\citet{van-paradijs96} carried out this test for the first time and showed that the DIM could indeed reproduce the transient-persistent dichotomy in X-ray binaries (XRBs) as long as irradiation heating is taken into account. The X-ray emission coming from the inner regions can indeed heat the outer parts of the accretion disc above the hydrogen ionization temperature and therefore stabilise the flow. Evidence for the influence of irradiation comes mainly from the higher optical to X-ray luminosity ratio observed in XRBs compared to cataclysmic variables (CVs) which should be less affected by irradiation due to smaller discs and accretion luminosities \citep{van-paradijs94}. 

Despite the limited number of sources available at that time, the work of \citet{van-paradijs96} represents one of the most fundamental test of the applicability of the DIM
to LMXBs. Fifteen years later, the number of known XRBs and the amount of data available to study them have increased significantly. 
We thus believe that it is time to revisit and refine this test, which is the aim of this paper. We also point out the recent work by \citet*{janiuk11} which discusses theoretical
aspects of the hydrogen ionization and radiation pressure instabilities and present  observational constraints  using a sample of Galactic black hole binaries. 

This paper is organised as follows. In Sect. 2, we detail the main assumptions and equations leading to the relation between the critical mass accretion rate and the orbital period on which the test rely. 
Sect. 3 presents our sample of sources and describes the data analysis and the method we used to estimate the mass transfer rate from the companion star. 
We present the results in Sect. 4 and discuss the outcomes in Sect. 5. The last section summarises and concludes this work.

\section{A simple test of the disc instability model}

An accretion disc is globally stable, and does not experience outbursts, whenever all its annuli are stable. The disc can be in a hot, stable
equilibrium if the local accretion rate $\mdot(R)$ is higher than the critical accretion rate of the hot, ionized state, $\mcrit(R)$, for all radii $R$ in
the disc. Since $\mcrit(R)$ increases with radius, this condition can be fulfilled if the mass transfer rate from the companion star, $\mext$, 
is higher than the the critical accretion rate at the outer radius of the disc $\mcrit(R_{\rm out})$. This is the standard scenario for persistent X-ray binaries. 

However if  $ \mext < \mcrit(R_{\rm out})$, the disc enters a limit cycle behaviour, oscillating between the ionized state (outburst) and the neutral state (quiescence),
i.e. the system is transient.

When the effect of irradiation is neglected, the critical mass accretion rate above which a ring of matter located at a radius $R$ is in the stable hot state 
can be expressed as follows \citep*[][appendix A]{lasota08}: 

\begin{equation}
\label{mcrit-nonirr}
\dot{M}_{\rm crit}= 8.07\times10^{15}~\alpha_{0.1}^{-0.01}~R_{10}^{2.64}~M_1^{-0.89} \gs
\end{equation}
where $M_1$ is the mass of the compact object in solar units, $\alpha = 0.1 \, \alpha_{0.1}$ is the viscosity parameter and $R = R_{10} \, 10^{10} \rm{cm}$.
When irradiation is taken into account, the critical accretion rate becomes  \citep*[][appendix A]{lasota08}:
\begin{equation}
\label{mcrit-irr}
\begin{aligned}
\dot{M}_{\rm crit} & = 9.5 \times 10^{14} 
                    ~{\cal C}_{-3}^{-0.36}
                    ~\alpha_{0.1}^{ 0.04+ 0.01\log{\cal C}_{-3}}
                    ~R_{10}^{2.39-0.10\log{\cal C}_{-3}} \\
                    &  \times   ~M_1^{-0.64+ 0.08\log{\cal C}_{-3}}\,\rm g\,s^{-1}
\end{aligned}
\end{equation}
where $\calc = 10^{-3} \,  \calc_{\rm -3}$ is the irradiation constant defined by the irradiation flux:

\begin{equation}
\sigma T_{\rm irr}^4  = \calc \frac{\dot{M} c^2}{4 \pi R^2}
\end{equation}

 The above expression of $\dot{M}_{\rm crit}$ has been derived by investigating the local vertical structure of an irradiated disc where X-ray heating is limited to a thin layer above the disc and appears in the boundary condition at the surface \citep[see][for details]{dubus99}:
\begin{equation}
T_{\rm surf}^4 = T_{\rm eff}^4 + T_{\rm irr}^4
\end{equation}
where $T_{\rm surf}$ is the surface temperature and $T_{\rm eff}$ the effective temperature in the absence of irradiation. 

As mentioned previously, the overall disc will be stable if the mass transfer rate from the donor star $\dot{M}_{\rm ext}$ is higher than the critical accretion rate (Eq. \ref{mcrit-nonirr} or \ref{mcrit-irr}) at the outer radius $R_{\rm out}$. For a given mass ratio $q=M_2/M_1$, one can compute the ratio $R_{\rm out}/a$ between the outer radius and the orbital separation following the calculations given in e.g. \citet{paczynski77}. The orbital separation can then be written as a function of the orbital period $P$ (in hr) using Kepler's law:
\begin{equation}
\label{a}
a =  3.53 \times 10^{10} M_1^{1/3} \left( 1+q \right)^{1/3} P_{\rm hr}^{2/3} \; \rm{cm} 
\end{equation}
Consequently, for a given set of parameters $\alpha$, $\calc$, $M_1$ and $q$, we can derive a relation between the critical accretion rate at the outer radius of the disc and the orbital period. 

The test therefore consists in estimating the mass transfer rate of a sample of persistent and transient XRBs (black holes and neutron stars) with known orbital periods and compare their position in a $\dot{M}-P$ diagram with the transient/persistent separation predicted by the $\dot{M}_{\rm crit}(P)$ relation derived from the DIM.

Note that the irradiated case (Eq. \ref{mcrit-irr}) contains a free parameter, the irradiation constant $\calc$. The latter is a measure of the fraction of the X-ray luminosity
intercepted and thermalised in the upper layers of the outer disc and, as such, contains information on the irradiation geometry, radiative efficiency, X-ray albedo, X-ray spectrum etc. It is thus a 
 poorly constrained quantity. Its constancy can also be questioned. \citet*{esin00} and \citet{esin00a} present some evidence of changes of $\calc$ during the outbursts of GRO J1655-40, A0620-00 and GRS 1124-68. The recent results on accretion disc winds in black hole XRBs from \citet{ponti12} also suggest a significant variation of the influence of irradiation during outbursts. Nonetheless, for simplicity, we restricted ourselves to constant $\calc$ in this study and started by assuming a value of $10^{-3}$ found to be compatible with the observed optical magnitudes and stability properties of persistent X-ray binaries \citep{dubus99}. This value is also consistent with the results from, e.g., \citet{de-Jong96} or \citet{gierlinski09} who found $\calc \sim 10^{-2} - 10^{-3}$ by comparing
 irradiated disc models to optical/UV and X-ray data.

\section{Data selection and analysis}\label{sec2}
 
To test the validity of the stability criteria (Eq. \ref{mcrit-nonirr} and \ref{mcrit-irr}), we selected a sample of 52 XRBs with known orbital periods, reasonably well-sampled light curves and for which we think that accretion proceeds via a disc. Twenty three of these contain a black hole (BH) or a black hole candidate (BHC) and the other 29 host a neutron star (NS). This is more than twice the number of sources used in the original work by \citet{van-paradijs96}. We classified the sources as BH or NS according to the estimates of the mass of the compact object and the detection of nuclear bursts. If the BH nature of the compact object is only suggested by the global spectral and temporal properties of the source, we classified it as a BHC. However, the exact nature of the compact object has a little influence on the results of the test since  the $\dot{M}_{\rm crit}(P)$ relation derived from the DIM is very weakly dependant on the primary mass (see Sect. \ref{discussion}).

All the sources in our sample are considered as low-mass XRBs except the three BH systems Cyg X-1, LMC X-1 and LMC X-3 and the three NS systems SMC X-1, LMC X-4 and Cen X-3. These objects likely host a high-mass companion star producing winds that are later accreted by the compact object. However, the X-ray spectro-temporal properties of these sources indicate the presence of an accretion disc fed by these focused winds. In some cases (e.g. Cen X-3), it is very likely that the companion star fills its Roche lobe and therefore transfers mass through the L1 point in addition to the winds. We therefore included them in our sample taking into account that the size of the disc derived from the orbital period might be overestimated.

The BH and NS samples contain respectively 6 and 17 sources that we classified as persistent systems. To do so, we assumed that the X-ray sky is well covered ($\sim 80 - 90 \%$) since 1970 \citep{chen97}. If the source was discovered in the early 1970s or before and is active since then, we classified it as persistent. By ``active'', we mean here that it 
never went below the detection threshold of the instruments that observed it throughout the years.

Note that two sources do not easily fit into this simplistic bimodal classification.
The BHCs and microquasars 1E 1740.7-2942 and GRS 1758-258 have been discovered respectively by the \textit{Einstein} satellite in 1984 \citep{hertz84} and the \textit{Granat} satellite in 1990 \citep{mandrou90}. Both sources are commonly considered as persistent in the literature since they remained active since their discovery. No X-ray source was identified at the location of 1E 1740.7-2942 during previous observations of the Galactic centre region by the satellites \textit{Uhuru, Ariel V} and \textit{SAS 3} \citep[see][and references therein]{pavlinsky92}. However, given the variability of the source  and the sensitivity and angular resolution of these instruments, we cannot exclude that 1E 1740.7-2942 was already active but not detected. 

In the case of GRS 1758-258, due to its proximity ($\sim$ 40 arcmin) to the persistent neutron star Z-source GX 5-1, it is very  likely that both sources were considered the same before \textit{Granat} was able to identify them separately (\citealt{levine84} already suspected that the hard X-ray flux assigned previously to GX 5-1 may originate from another nearby X-ray source). Given that an X-ray source is already present in the second \textit{Uhuru} catalog \citep{giacconi72} at a location consistent with GRS 1758-258, it is possible that the source was already active before its identification. Bearing all these considerations in mind, at first instance, we will treat these sources as persistent

We also stress that among the 6 persistent BH systems, 3 of them are the high-mass XRBs mentioned previously and 2 of them are the black hole candidate 1E 1740.7-2942 and GRS 1758-258 
for which, as we just warned, the persistent nature is unclear. The last one, 4U 1957+115, agrees well with our persistency criterion. The source remained persistently in soft states since its discovery \citep{wijnands02} and recently provided very deep radio upper limits supporting the jet quenching paradigm in soft states black hole XRBs \citep{russell11}. The X-ray properties of the source strongly suggest the system is hosting a black hole \citep[see e.g.][]{nowak08} but there is no firm evidence on the nature of the compact object.
The persistent BH sample should thus be considered with caution.
 
Note that the proportion of sources with known orbital periods corresponds to $\sim 50 \%$ of the known sources in the case of BH and about $30 \%$ in the case of NS systems.
These proportions must be kept in mind with respect to the results of this study. 
The sample we have selected might also be biased toward high inclination sources for which it is easier to determine the orbital period. 

For each source we selected, we list in Table \ref{tab-source-bh} and \ref{tab-source-ns} the mass, distance and orbital period we used for this study.
 
 \begin{table*}
 \centering
 \caption{\sc Black Holes X-ray Binaries Parameters}
\label{tab-source-bh}
\begin{tabular}{lcccccr}
\hline
\hline
Source & Type & $M$(M$_{\odot}$) & $D$(kpc) & $P\rm{_{orb}}$(hr) & Notes\\
\hline
GRO~J0422+32 & T &  3.97$\pm$0.95 & 2.75 $\pm $0.25 & 5.092 &  \\
A~0620-00 & T & 11.0$\pm$1.9 & 1.2 $\pm$ 0.4 & 7.75& \\
GRS~1009-45 & T  & $>3.9$ & 5.7 $\pm$ 0.7 & 6.84 & \\
XTE~J1118+480 & T& 8.53 & 1.8 $\pm$ 0.6 & 4.08 &  \\
GS~1124-684 & T & 4-11 & 5.5 $\pm$ 1.0 & 10.38\\
GS~1354-64 & T & $>5.75$ & $\ge$25 & 61.07& (1)\\
4U~1543-47 & T & 8.4-10.4 & $7.5 \pm 0.5$ & 26.8\\
XTE~J1550-564 & T & 10.5$\pm$1.0& $5.3 \pm 2.3$ & 37.25\\
XTE~J1650-500 & T & $>2.73$ & $2.6 \pm 0.7$ & 7.63 & (2)\\
GRO~J1655-40 & T & 6.3$\pm$0.5 & $3.2 \pm 0.2$& 62.88& (3)\\
MAXI J1659-352 & T, BHC & 3.6-8.0 & 8.6 & 2.4 & (4)\\
GX~339-4 & T & $>5.3$ & $\sim$8 & 42.14 & (5)\\
4U~1705-250 & T & $> 4.73$ & $8.6 \pm 2.0$ & 12.54\\
Swift J1753.5-0127& T, BHC & $ ... $ & $\sim$8 & 3.2 & (6) \\
GRS~1915+105 & T & 14$\pm$4 & $9.0 \pm 3.0$ & 739.2 & (7)\\
GS~2000+25 & T & 4.8-14.4 & $2.7 \pm 0.7$ & 8.26\\
V404 Cyg (GS~2023+338) & T & 12$\pm$2 & $2.39 \pm 0.14$ & 155.4 & (8)\\
\\
1E 1740.7-2942 & P?, BHC & ... & $\sim$8.5 & 305.5 & (9)\\
GRS 1758-258 & P?, BHC & ... & $\sim$8.5  & 442.8 & (9)\\
\\
4U 1957+115 & P, BHC & ... & $\ge 7$ & 9.33 & (10) \\
Cyg X-1& P, HMXB & $14.8 \pm 1.0$ & $1.86 \pm 0.12$ & 134.4 & (11)  \\
LMC X-1 & P, HMXB & $10.91 \pm 1.41$ & $48.10 \pm 2.22$ & 93.6 & (12)\\
LMC X-3 & P, HMXB & $11.1 \pm 1.4$ & $52.0 \pm 0.6$ & 40.8 & (13) \\
\hline
\end{tabular}
\begin{quote}
Col. (Source): the source name. (Type): ``T'' and ``P'' indicate transient and persistent systems respectively. ``HMXB'' indicates a high mass X-ray binaries.
($M$): the mass of the compact object.
($D$): the distance to the source. ($P_{\rm orb}$): the orbital period of the system.
\newline Unless otherwise stated, parameters are from \citet*{liu05a,liu06,liu07}  and \citet*{jonker04}.
\newline \textsc{Notes:} (1) Lower limit on the distance from \citet{casares09}. (2) Distance from \citet{homan06}.
(3) Note that \citet{foellmi09} argues for an upper limit to the distance of 2 kpc. 
(4) Parameters from \citet{kuulkers12} and \citet{yamaoka12}
(5) Distance from \citet{zdziarski04}.
(6) Parameters from \citet{zurita08} and \citet{cadolle-bel07}. 
(7) Orbital period from \citet*{neil07} and distance from \citet{fender99} and \citet*{chapuis04}.
(8) Distance from \citet{miller-jones09a}.
(9) No estimate of the mass of the compact object. We adopted a Galactic center distance for the source. Orbital period from \citet{smith02a}.
(10) Distance from \citet{yao08} and \citet{nowak08} (see also \citealt{nowak12}), orbital period from \citet{thorstensen87} and \citet{bayless11}.
(11) Mass from \citet{orosz11}, distance from \citet{reid11} and orbital period from \citet{brocksopp99}.
(12) Parameters from \citet{orosz09}.
(13) Mass from \citet{gierlinski01}, distance from \citet{di-benedetto97} and orbital period from \citet{hutchings03}.
\end{quote}
\end{table*}

 \begin{table*}
 \centering
 \caption{\sc Neutron Star X-ray Binaries Parameters}
\label{tab-source-ns}
\begin{tabular}{lcccccr}
\hline
\hline
Source & Type & $M$(M$_{\odot}$) & $D$(kpc) & $P\rm{_{orb}}$(hr) & Notes\\
\hline 
IGR~J00291+5934 & T & ... & 2.6-3.6 & 2.46\\
EXO~0748-676 & T & ... & 5.9-7.7 & 3.82\\
Cen~X-4 & T & 1.3$\pm$0.6 & $1.2 \pm 0.2$ & 15.10\\
4U~1608-52 & T &  ... & 2.8-3.8 & 12.89\\
XTE J1710-281 & T & ... & 12-16& 3.28 \\
1A~1744-361 & T & ... & $<$9 & 1.62\\
GRS 1747-312 & T & ... & 9.5 $\pm$ 3.0 & 12.36\\
XTE~J1751-305 & T & ... & $\sim$8.5 & 0.71\\
SAX~J1808.4-3658 & T & $<$2.27 & 3.4-3.6 & 2.01\\
XTE~J1814-338 & T & ... & 8.0$\pm$1.6 & 4.27\\
Aql~X-1 & T & ... & 4.4-5.9 & 18.95\\
XTE~J2123-058 & T & 1.5$\pm$0.3 & 8.5$\pm$2.5& 5.96\\
\\
LMC X-2 & P & ... & 50 $\pm$ 2 & 8.16 &\\ 
4U 0614+091 & P & ... & 2-3 & 0.8 & (1) \\
4U 1323-62 & P & ... & $\sim$ 10 & 2.9 \\
Sco X-1 & P & ... & 2.8$\pm$0.3& 18.9 \\
4U 1636-536 & P & ... & 3.7-4.9 & 3.8 \\
GX 349+2 & P & ... & $5.0 \pm 1.5$ & 22.5 & (2) \\
Her X-1 & P & 1.1 $\pm$ 0.4 & 6.6 $\pm 0.4$ & 40.8 & (3) \\
4U 1735-444 & P & $> 0.1$ & 8.0-10.8 & 4.6 \\
4U 1746-370 & P & ... & 11.0 & 5.16 \\
4U 1820-303 & P & ... & 7.6$\pm$0.4& 0.19 & (4)\\
4U 1850-087 & P & ... & 8.2$\pm$0.6& 0.34& (4) \\
4U 1916-053 & P & ... & 8.9$\pm$1.3& 0.83 \\
4U 2129+12 (AC 211) & P & ... & $10.3 \pm 0.4$ & 5.96& (4) \\
Cyg X-2 & P & $1.78 \pm 0.23$ & $7.2 \pm 1.1$ & 236.2 \\
\\
SMC X-1 & P, HMXB & 1.04 $\pm$ 0.08 & 60 & 93.36 & (3) \\ 
LMC X-4 & P, HMXB & 1.29 $\pm$ 0.05  & 50 & 33.79 & (3) \\ 
Cen X-3 & P, HMXB & 1.49 $\pm$ 0.08  & 9 $\pm$ 1 & 50.16 & (3) \\ 
\hline
\end{tabular}
\begin{quote}
Col. (Source): the source name. (Type): ``T'' and ``P'' indicate transient and persistent systems respectively. ``HMXB'' indicates a high mass X-ray binaries.
($M$): the mass of the compact object.
($D$): the distance to the source. ($P_{\rm orb}$): the orbital period of the system.
\newline Unless otherwise stated, parameters are from \citet*{liu05a,liu06,liu07}  and \citet*{jonker04}.
\newline \textsc{Notes:} 
(1) Orbital period from \citet{int-zand07}.
(2) Distance from \citet{christian97}.
(3) Mass from \citet{rawls11}.
(4) Distance from \citet{kuulkers03}.
\end{quote}
\end{table*}

\subsection{Estimating the mass transfer rate}

The bulk of the work presented here was to estimate, for each source, the mass transfer rate from the donor star.
For persistent sources, this task is fairly straightforward. If we assume that the accretion rate in the disc, averaged over time, is equal to \mext , 
we can estimate the latter by measuring the average X-ray luminosity:
\begin{equation}
L_{\rm X} = \eta \mext c^2
\end{equation}
where $c$ is the speed of light and $\eta$ is the radiative efficiency of accretion.
 
For transient systems, the standard procedure to estimate \mext\ is based on the assumption that the systems mainly store mass during quiescence
and accrete it suddenly during an outburst.  The rate at which the mass accumulates in the disc, $\dot{M}_{\rm accum}$, can be estimated from the integrated
 X-ray luminosity from the system during an outburst, $\Delta E$,  and from the outburst recurrence time, $\Delta t$:

\begin{equation}
 \dot{M}_{\rm accum} = \frac{\Delta E}{\Delta t  \eta  c^2}
\end{equation}
If the mass accretion rate during quiescence is negligible compared to the mass transfer rate from the secondary, we can then use $ \dot{M}_{\rm accum}$ as a proxy
for $\mdot_{\rm ext}$.  This assumption ignores the possibility of a highly inefficient accretion flow which could hide a substantial accretion of mass in quiescence \citep*{narayan97,menou99}.
Furthermore, estimating $\dot{M}_{\rm accum}$ from X-ray emission alone does not take into account mass loss through winds and jets \citep*[e.g.,][]{fender03}. 
Consequently, for both reasons, our estimates of $\mdot_{\rm ext}$ should be considered as lower limits to the actual mass transfer rate.

For transient BH systems, the radiative efficiency may vary during an outburst. It is generally thought that the accretion flow is radiatively
inefficient during faint hard state with $\eta \propto \mdot$ while it is radiatively efficient during soft states with $\eta \sim 0.1$. In addition, the total
X-ray luminosity usually remains relatively constant \citep[within a factor of $\sim$ 2, e.g.,][]{zhang97} during the hard to soft state transition which does not suggest a drastic change in the
radiative efficiency. Therefore, during the bright hard state phases, $\eta$ should not be significantly lower than 0.1.
Consequently, we adopted the following prescription for $\eta$ when deriving $\mext$ for a transient BH XRB:
 \begin{equation}
\label{etabh}
\begin{aligned}
\eta & = 0.1 \left( \frac{\mdot}{0.01 \medd }\right) \; ; & \quad \quad {\rm for}\; \; L_{X} < 1 \% \ledd \\
\eta & = 0.1 \; ; & \quad \quad {\rm for}\; \; L_X > 1 \% \ledd
\end{aligned}
\end{equation}
where the Eddington accretion rate is defined by $\ledd = 0.1 \medd  c^2$. 
Accretion onto neutron stars is assumed to be radiatively efficient due to the presence of a solid surface on which the accreted material can eventually radiate 
 its remaining gravitational energy. We thus used $\eta  = {\rm const} =  0.1$ to derive $\mext$ for NS systems.
Note that we have chosen 0.1 as the maximum value for $\eta$, although the radiative efficiency can reach higher values for matter
rotating around highly spinning black holes. However, $\eta$ departs significantly from 0.1 only for very high spin values (e.g., $\eta \sim 0.2$ for
a spin of 0.9, \citealt{bardeen72} ), therefore, this shouldn't affect our results significantly for most of our sources.

To obtain reliable measures of $\Delta E$, we used all observations of the binaries we selected which were publicly available in the archive 
of the \textit{Rossi X-ray Timing Explorer} (\rxte) satellite. 
The data reduction and model fitting scheme is detailed in the next section. Some objects we selected
have been inactive since the launch of \rxte\ in December 1995. We thus used estimates of either $\Delta E$ or \mext\ found in the literature that we
corrected for updated distance if necessary. 
From model fitting to the X-ray spectra, we produced X-ray light-curves in the 0.1-200 keV energy band for each sources.  
Then, we fit the light-curves with combinations of Gaussians  or polynomial functions to integrate the luminosity and derive the energy liberated during each outburst. 
 In the case where an outburst was covered by the All Sky Monitor (ASM) but only partially covered by the Proportional Counter Array (PCA) pointed observations (typically, the rising phase was missing), we extrapolated 
 the function used to fit the PCA lightcurve in order to match the shape of the ASM lightcurve. If the outburst was entirely missed by the PCA, we estimated the total energy liberated
 using the ASM lightcurve and the conversion of 1 ASM count/s to $7.7 \times 10^{-10} \erg$ proposed by \citet{int-zand07}.  

Some objects underwent several outbursts which allows for a precise estimate of the recurrence time. Some other entered only once into an active phase. In that case we used a lower limit to $\Delta t$ (leading to an upper limit to \mext ) by considering that the X-ray sky is well covered since 1970 \citep{chen97}. Therefore, $\Delta t > \rm{max}(\Delta t_{1970}, \Delta t_{2012})$, where $\Delta t_{1970}$ and $\Delta t_{2012}$ are the times elapsed from 1970 to the outburst start date and from the outburst to 2012 respectively.

\subsection{\rxte\ data reduction}

We reduced \rxte\ data using the {\tt HEASOFT} software package version 6.11 following the standard procedure described in the \rxte\ cookbook\footnote{http://heasarc.gsfc.nasa.gov/docs/xte/data\_analysis.html}.

We extracted spectra from the Proportional Counter Array (PCA; \citealt{jahoda06}) for which we only used the Proportional Counter Unit (PCU) 2. This is the only operational unit across all the mission and is the best-calibrated detector out of the five PCUs. A systematic error of 1 per cent was added to all channels.
We obtained hard X-ray spectra from the High Energy Timing Experiment (HEXTE) using both cluster A and B until December 2005. After this date, due to problems in the rocking motion of Cluster A, we extracted spectra from Cluster B only. On the 14th of December 2009, Cluster B stopped rocking as well. From this date we thus used only PCA data in our analysis.  

\subsection{Model fitting}

We performed simultaneous fits to the PCA and HEXTE spectra in \textsc{Xspec} version 12.7.0 \citep{arnaud96} using a floating normalisation constant to allow for cross-calibration uncertainties. Channels out of the $3-50$ keV range for PCA spectra and out of the $20-200$ keV range for HEXTE spectra were ignored from the fit.

The main objective of the X-ray spectral analysis was to obtain a reasonable estimate of the bolometric luminosity. 
We have chosen the 0.1-200 keV band as a proxy for the bolometric luminosity. Note that the \rxte\ data do not constrain 
the models outside of the 3-200 keV range. The 0.1-200 keV flux is thus obtained by extrapolating the best-fitting models 
using an energy response matrix extended down to 0.1 keV (using the {\tt extend} command in \textsc{Xspec}).

By choosing the 0.1-200 keV flux as a proxy for the bolometric flux, we are neglecting a fraction of the accretion luminosity.
To estimate the importance of this missing fraction, we simulated with \textsc{Xspec} an irradiated accretion disc with a comptonisation component using the \textsc{diskir} model \citep{gierlinski08,gierlinski09}. Very briefly, this model is an extension of the \textsc{diskbb} model that includes comptonisation of the blackbody disc's photons by a corona of hot electrons \citep[based on \textsc{nthcomp}][]{zdziarski96,zycki99}, as well as disc irradiation of both the inner and outer regions of the disc. 
We first simulated typical soft states and hard states spectra of BHXRBs using the parameters given in \citet{gierlinski09}. We then calculated the ratio between the 0.1-200 keV flux and the $10^{-6}-500$ keV flux (chosen to represent the bolometric flux). We found that, in the less favourable case, the 0.1-200 keV flux represents 75 \% of the bolometric flux.

However, since we fit the data above 3 keV only, we also need to consider the case where the disc was missed because its contribution above 3 keV was not significant enough (typically when $kT_{\rm disc} < 0.4$ keV). Extrapolating our best fitting model down to 0.1 keV implies that we only considered the contribution of the comptonisation component in the 0.1-3 keV band and therefore probably missed a significant fraction of the bolometric flux. To estimate the fraction we missed in these particular cases,  we simulated a hard state spectrum using \textsc{diskir} with $kT_{\rm disc}  = 0.4$ keV and  measured the bolometric flux. We then compared the latter with the 0.1-200 keV flux obtained
using a simple (powerlaw + high energy cutoff) model with the same normalisation and spectral parameters than the comptonisation component of the \textsc{diskir} spectrum. We found that
the 0.1-200 keV in that case represents  $\sim 50 \%$ of the bolometric flux. Consequently, during observations where the disc's inner temperature is low, we might underestimate the bolometric flux by a factor of 2. However, these situations occur essentially during the low luminosity states.  For the purpose of estimating the mass transfer rate, we are mainly interested in the high luminosity states where most of the mass is accreted. Therefore, our results shouldn't be affected significantly.

Significant variation of the hydrogen column density may occur during outburst of transient XRBs (see e.g. \citealt{cabanac09}, but also \citealt{miller09}). However, the PCA response falls below 2 keV and doesn't allow the interstellar absorption to be constrained properly. We thus fixed the \nh\ to the commonly accepted values for the binaries. We used the photoelectric absorption model \textsc{phabs} using abundances of \citet*{wilms00} and  cross-sections of \citet{balucinska-church92}.

Given the high number of spectra to model, we automated the fitting procedure. For BH sources, we followed the method described in \citet{dunn10} for which we refer the reader 
for full details. For NS systems, we followed \citet*{lin07} and adopted their hybrid model (blackbody + broken powerlaw for hard states and disc blackbody + blackbody + constrained broken powerlaw for soft states) to fit the spectra. A full description of the model and its performance against several desirability criteria (e.g. $L_{X} \propto T^4$) are given in 
\citet*{lin07}.

Table \ref{bh-mdot} and \ref{ns-mdot} present our estimates of the mass transfer rate and the average outburst recurrence time (if relevant) for our BH and NS samples respectively.

 \begin{table}
 \centering
 \caption{\sc Black Holes X-ray Binaries}
\label{bh-mdot}
\begin{tabular}{lccc}
\hline 
\hline 
Source & \mext\ & $\Delta t_{\rm av}$ & $ \mext / \dot{M}_{\rm crit}$\\
& ($ {\rm g \, s}^{-1}$) & (yr) &\\
\hline
GRO J0422+32 & $< 1.7 \times 10^{15}$ (1) & $> 22$ & $ < 2.6 \times 10^{-2}$  \\
A0620-00 & $3.4 \times 10^{15}$ (1) & 58 & $ 2.7 \times 10^{-2}$ \\
GRS~1009-45 & $ < 1.7 \times 10^{16}$ (1) & $> 23$ & $ < 0.16$ \\
XTE J1118+480 & $1.3 \times 10^{16}$ & 5 &  $0.28$ \\
GS 1124-684 & $< 2.1 \times 10^{16}$ (1) & $> 21$& $ < 0.10$ \\
GS 1354-64 & $> 1.0\times 10^{17}$ & 10 & $ > 2.9 \times 10^{-2} $ \\
4U 1543-47 & $7.1\times 10^{16}$ & 10 & $7.7 \times 10^{-2}$ \\
XTE J1550-564 & $1.0\times 10^{17}$ & 20 & $6.5 \times 10^{-2}$ \\
XTE J1650-500 & $<6.9\times 10^{15}$ & $> 31$ & $< 5.5 \times 10^{-2}$ \\
GRO J1655-40 & $5.5\times 10^{16}$ & 10 & $1.6 \times 10^{-2}$ \\
MAXI J1659-352 & $8.9\times 10^{15}$ & $> 40$ & $< 0.44$ \\
GX 339-4 & $8.0\times 10^{17}$ & 2 & $0.43$ \\
4U 1705-250 & $< 1.8\times 10^{16}$ (1) & $> 24$ & $< 6.5 \times 10^{-2}$\\
Swift J1753.5-0127 & $< 7.2 \times 10^{16}$ & $> 35$ & $< 2.3$ \\
GRS 1915+105 & $ < 9.8 \times 10^{18}$ & $> 22$ & $< 5.5 \times 10^{-2}$ \\
GS 2000+25 & $< 4.1\times 10^{15}$ (1) & $> 23$ & $< 2.9 \times 10^{-2}$ \\
V404 Cyg & $ 5.9 \times 10^{16}$ (1) & $ 25 (2) $ & $ 3.9 \times 10^{-3}$ \\
\hline
1E 1740.7-2942 & $3.4 \times 10^{18}$ & -- & $7.7 \times 10^{-2}$ \\ 
GRS 1758-258 & $4.2 \times 10^{18}$ & -- & $5.3 \times 10^{-2}$ \\
4U 1957+115 & $ > 9.2 \times 10^{16}$ & -- & $> 0.54$ \\
Cyg X-1 & $8.9 \times 10^{17}$ & -- & $ < 7.5 \times 10^{-2}$  \\
LMC X-1 & $4.1 \times 10^{18}$ & -- & $< 0.61$ \\
LMC X-3 & $4.2 \times 10^{18}$ & -- & $< 2.34$ \\
\hline
\end{tabular}
\begin{quote} 
(1) Estimated from outburst fluences given in \citet{chen97}
\newline (2) Recurrence time estimated from \citet{richter89} 
\end{quote}
\end{table}

\begin{table}
 \centering
 \caption{\sc Neutron Stars X-ray Binaries}
\label{ns-mdot}
\begin{tabular}{lccc}
\hline
\hline
Source & \mext\ & $\Delta t_{\rm av}$ & $ \mext / \dot{M}_{\rm crit}  $\\
& ($ {\rm g \, s}^{-1}$) & (yr) & \\
\hline
IGR~J00291+5934 & $3.2 \times 10^{14}$ & 3 & $ 2.3 \times 10^{-2}$ \\
EXO~0748-676 & $ < 2.8 \times 10^{16}$ & $> 28$ & $ < 1.0$\\
Cen~X-4 & $ 2.4 \times 10^{15}$ (1)& 10 & $1.0 \times 10^{-2}$  \\
4U~1608-52 & $ 6.1 \times 10^{16}$ & 0.56 & $0.33$ \\
XTE J1710-281 & $<1.0\times 10^{16}$ & $>28$ & $<0.86$ \\
1A~1744-361 & $ < 1.1 \times 10^{16}$ & 2 (2) & $ < 1.5$  \\
GRS 1747-312 & $7.1 \times 10^{16}$ & 0.37 & 0.40  \\
XTE~J1751-305 & $3.8 \times 10^{14}$ & 3.8 (3) & $ 0.20$  \\
SAX~J1808.4-3658 & $1.1 \times 10^{15}$ & 2 & 0.11  \\
XTE~J1814-338 & $ < 3.8 \times 10^{14}$ & $> 33$  & $ < 1.2 \times 10^{-2}$ \\
Aql~X-1 & $3.8\times 10^{16}$ & 0.55 & 0.11 \\
XTE~J2123-058 & $ < 4.3 \times 10^{14}$ & $> 28$  & $ < 7.8 \times 10^{-3}$ \\
\hline
LMC X-2 & $2.4\times 10^{18}$ & ... & $ 27 $ \\
4U 0614+091 & $2.9\times 10^{16}$ & ... & $ 13 $ \\
4U 1323-62 & $6.1\times 10^{16}$ & ... & $ 3.5 $ \\
Sco X-1 & $1.8\times 10^{18}$ & ... & $ 5.2$  \\
4U 1636-536 & $7.9\times 10^{16}$ & ... & $ 2.9$ \\
GX 349+2  & $1.5\times 10^{18}$ & ... & $ 3.4 $\\
Her X-1 & $8.0\times 10^{17}$ & ... & $ 0.68 $ \\
4U 1735-444 & $4.0\times 10^{17}$ & ... & $ 11$ \\
4U 1746-370 & $2.6\times 10^{17}$ & ... & $ 6.0$ \\
4U 1820-303 & $7.9\times 10^{16}$ & ... & $ 3.4 \times 10^{2}$ \\
4U 1850-087 & $2.4\times 10^{16}$ & ... & $ 41 $\\
4U 1916-053 & $8.2\times 10^{16}$ & ... & $ 34 $ \\
4U 2129+12 & $2.5\times 10^{17}$ & ... & $ 4.6 $  \\
Cyg X-2  & $1.9\times 10^{18}$ & ... & $ 0.1 $ \\
SMC X-1 & $5.5\times 10^{18}$ & ... & $ 1.3 $ \\
LMC X-4 & $6.7\times 10^{18}$ & ... & $ 7.7 $ \\
Cen X-3 & $9.4\times 10^{17}$ & ... & $ 0.58 $ \\
\hline
\end{tabular}
\begin{quote}
(1) Estimated from outburst fluences given in \citet{chen97}
(2) Recurrence time and accretion rate estimated from the activity of the source since 2003.
(3) From \citet{markwardt02}.
\end{quote}
\end{table}

\subsection{Dominant sources of uncertainty in the estimate of $\mext$}

The value of \mext\ we derive using the method described above will of course depend on the estimate of the distance to the source. 
If we assume an error of $25 \%$ on the distance, this will introduce an uncertainty of a factor of 2 on the estimate of \mext.
However, one can calculate using Tables \ref{tab-source-bh} and \ref{tab-source-ns} that the average error on the distance for the sources of our sample is $\sim 20 \%$ for the black holes 
and $\sim 15 \%$ for the neutron stars. Therefore, an uncertainty of a factor of 2 on the value of \mext\ should be representative. 

Owing to the smaller projected area of the disk, edge-on systems appear less luminous than face-one systems for the same intrinsic luminosity.
The observed apparent X-ray luminosity is reduced by a factor of $\sim \cos i$, where $i$ is the angle between the line of sight and the rotation axis of the disc.
Inclination could also affect the observed X-ray luminosity  if e.g. the disc starts to become geometrically thick and self-screens part of its inner regions, and/or 
if an equatorial wind from the disc becomes optically thick, so that the luminosity seen at high inclinations is lower than the intrinsic flux seen at low inclinations. 
If these effects are significant, the mass transfer rate we infer from the total X-ray luminosity would be underestimated. 

Errors introduced by the model fitting procedure are not easy to estimate given that we extrapolate the models out of the energy range covered by the data.
After running several tests comparing fluxes obtained using extrapolated models from \rxte\ data only and fluxes obtained with models constrained at low energies
by \textit{Swift} data, we estimated that a typical error of a factor of 2 on the outburst fluences (flux integrated over time) introduced by the
fitting procedure should be reasonable.

An additional uncertainty is introduced by the efficiency factor $\eta$ with which gravitational energy is transformed to radiation and how representative the flux is for the mass accretion rate. However, the main contribution to the total energy liberated during an outburst arises mostly from the brightest phases where we assume $\eta =0.1$. This value is usually considered 
as an upper limit on the actual radiative efficiency for most of the sources, which means a lower limit on \mext. Therefore, this strengthen our previous caveat on our estimates of the mass transfer rate.

Estimating the mass transfer rate in transient systems partly relies on how precisely we are able to estimate the recurrence time, which in turn depends on how we define quiescence and outburst. As a basic definition, we considered that an outburst started (ended) when the source went above (below) the detection level of the ASM or the PCA depending on which one was the first (last) to detect it. Although very crude and instrument-dependent, this is the best definition we can use since regular monitoring of sources at very low luminosity is not yet possible. Since this definition of quiescence will systematically overestimate the inter-outburst time, this will lead to a lower limit on the actual mass transfer rate, strengthening again our previous caveat. For sources with decade-long quiescent period, an overestimate of the order of weeks or even month on the recurrence time will have a negligible impact on the estimate of \mext. For sources with short recurrence time ($\Delta t \la 2$ yrs, e.g., GX 339-4 for the black holes or Aql X-1 and 4U 1608-52 for the neutron stars), overestimating $\Delta t$ by $50 \%$ will underestimate $\mext$ by a factor 1.5.

A last point should be stressed regarding the recurrence time. Many examples of transient sources show inter-outburst times varying from 1 to more than 10 years for the same binary system which can lead to significant errors if one uses an averaged value of $\Delta t$.  Moreover, for several sources, the duration of an outburst is comparable or even greater than the time spent in quiescence (e.g. GX 339-4 since 2002). Since the total mass accreted during an outburst should correspond to the mass accumulated in the disc during quiescence plus the mass transferred to it during the outburst, neglecting the duration of the outburst in estimating $\Delta t$ can overestimate \mext\ by a factor of 2.
To minimise this, we calculated \mext\ for each individual outburst by dividing the total energy liberated
during the outburst by the time elapsed between the end of the previous outburst and the end of the studied one. 
We then averaged the values of \mext\ obtained for each individual outburst.

Table \ref{error} summarises the impact of the main sources of uncertainty on the estimates of the mass transfer rate. It appears from this table
that a symmetric error of a factor of 3 (quadratic sum of two factors of 2) should be taken into account but that, overall, $\mext$ is likely underestimated. 

\begin{table}
 \centering
 \caption{\sc Impact of the main sources of uncertainty}
\label{error}
\begin{tabular}{ll}
\hline
\hline
\bf Source of uncertainty  &\bf Impact on $\mathbf{\mext}$  \\
\hline
High quiescent $\mdot$ & Underestimated \\
Mass loss through winds and jets &  Underestimated \\
$25\%$ uncertainty on distance & Error of a factor 2 \\
Inclination effects &  Underestimated \\
Fitting procedure &  Error of a factor 2 \\
Radiative efficiency overestimated &  Underestimated \\
Recurrence time overestimated &  Underestimated \\
\hline
\end{tabular}
\begin{quote}
\end{quote}
\end{table}

\section{Results and Discussion}\label{discussion}

Figures \ref{dimns} and \ref{dimbh} plot the mass transfer rate against the orbital period for the NS and BH sources respectively, 
together with the predicted separation between transient and persistent systems in the irradiated and non-irradiated cases.
To plot the stability criteria we assumed $\alpha = 0.1$ and $\calc = 10^{-3}$. For the black hole systems, we assumed
a primary mass $M_1$ varying between $3 \msun$ and $15 \msun$ and for neutron stars we assumed $M_1 = 1.4 \msun$.
Finally, for both types of systems we considered a mass ratio $q$ varying between 0.1 and 1.
Using the above parameter space, the stability criteria can be written $\dot{M}_{\rm crit} = k P_{\rm hr}^b$ with, for the non irradiated case, $b = 1.76$, 
and $k = (2.6 \pm 0.9) \times 10^{16}$ for both the BH and the NS. In the irradiated case,
$b = 1.59$, $k= (3.9 \pm 1.6) \times 10^{15}$ for the BH and $k = (2.9 \pm 0.9) \times 10^{15}$ for the NS.
The stability limits are represented by the shaded grey areas on Fig.~\ref{dimns} and \ref{dimbh}.

\subsection{Neutron stars}

\begin{figure*}
	 \includegraphics[width=1\textwidth]{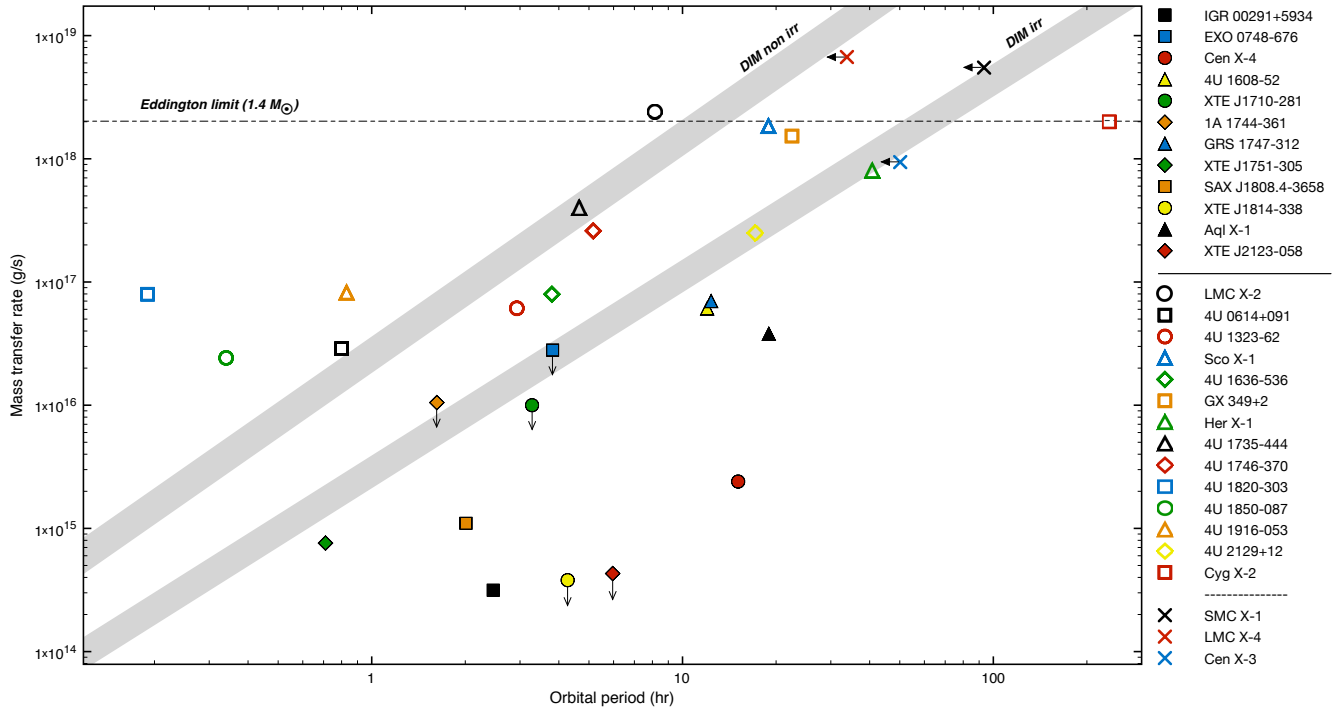} 
	\caption{Mass transfer rate as a function of the orbital period for XRBs with neutron stars. The transient and persistent LMXBs have been indicated with filled and open 
	symbols respectively, while the crosses indicate the high-mass, persistent systems. The shaded grey areas indicated `DIM irr' and `DIM non irr' represent the separation between persistent (above) and transient systems (below) according to the disc instability model when, respectively, irradiation is taken into account and when it is neglected. The horizontal dashed line indicates the Eddington accretion rate for a 1.4 \msun\ neutron star. The upper limits on the mass transfer rate are due to lower limits on the recurrence time except for 1A 1744-361 for which it arises from a upper limit on the distance to the source.
The three left closed arrows do not indicate actual upper limits on the orbital period of SMC X-1, LMC X-4 and Cen X-3. They emphasise that the radius of any accretion disk in these three high-mass XRBs is likely to be smaller than the one derived from the orbital period since they likely transfer mass by a (possibly focused) stellar wind instead of (or in addition to) fully developed Roche lobe overflow. In the legend, the solid horizontal line separates transient and persistent systems. The dashed horizontal line separates low-mass and high-mass persistent systems.
We can notice from this plot that the irradiated DIM seems to reproduce the transient-persistent division very well.}
	\label{dimns}
\end{figure*}

From a global standpoint, the neutron star data are in very good agreement with the theoretical predictions of the irradiated DIM. Indeed, we note that the stability line draws a fairly accurate limit between persistent and transient sources. Assuming that the main uncertainty in deriving the stability criterion lies in the irradiation constant $\calc$, we can use the narrow separation between the two groups of sources to roughly constrain its value. As we see on Fig. \ref{dimns}, for the value $\calc = 10^{-3}$ we initially assumed, the persistent sources 4U 2129+12 and Her X-1 lie already on the stability line. If we increase the irradiation constant, the line shifts toward lower mass transfer rate and for $\calc = 10^{-2}$ it reaches the transient systems 4U 1608-52, GRS 1747-312 and XTE J1751-305. Consequently, if the physical parameters assumed to derive the stability criterion are correct as well as our estimates of the mass transfer rates, the irradiation constant for NS XRBs should range between $10^{-2}$ and $10^{-3}$.

\subsubsection{Cyg X-2}
The only binary in our NS sample which does not agree with the DIM prediction is the persistent Z-source Cyg X-2 which should be transient according to its position in the diagram.  
However, the irradiated DIM predicts that the critical mass transfer rate corresponding to the large orbital period of Cyg X-2 should be an order of magnitude above the Eddington limit for a $1.4 \msun$ neutron star. Since we derived \mext\ from the (supposedly Eddington limited) X-ray luminosity, we might significantly underestimate the actual mass transfer at the outer edge of the disc. If the latter is truly an order of magnitude above \medd, either the X-ray emitting flow becomes radiatively inefficient or outflows carry away 90\% of the mass transferred from the donor star or the DIM has to be modified to properly describe super-Eddington accreting systems (for instance, to consider an Eddington limited irradiating flux). 

Another explanation would be that Cyg X-2 is in fact a transient source undergoing a very long outburst in a way similar to the well known black hole transient GRS 1915+105. Long orbital period systems imply large accretion discs which can potentially accumulate enough mass to fuel a decades long outburst.
To test this idea, we estimated the disc mass of Cyg X-2 before the onset of a hypothetical outburst using equation (8) of \citet{truss06} and the expression of the critical density 
of the cold neutral state ($\Sigma_{\rm crit}^{-}$) given in \citet[][appendix A]{lasota08}. We obtained\footnote{Note that, following \citet{truss06}, we also integrated the surface density over a portion of the disc comprised between $0.1 R_{\rm disc}$ and $R_{\rm disc}$ but we rather assumed an average surface density of $\Sigma_{\rm crit}^{-} (0.2 R_{\rm disc})$ instead of $\Sigma_{\rm crit}^{-}(0.1 R_{\rm disc})$ assumed by \citet{truss06}. We thought this value was slightly more consistent with the results of \citet{dubus01}} a disc mass $M_{\rm disc} = 2.5 \times 10^{28} g$. 
Assuming that the matter is accreted onto the compact object at the Eddington rate, such a disc should be able to sustain a $\ga 80$ years long outburst. Given that Cyg X-2 was discovered 46 years ago, it could indeed be a transient source. 

\subsubsection{Remarks}
It is also worth noting that the Z-sources of our sample (Sco X-1, Cyg X-2, LMC X-2 and GX 349+2) lie on or very close to the Eddington limit, in agreement with the known characteristic
of this class of neutron stars XRBs.
We note as well that the high-mass systems SMC X-1 and LMC X-4 are accreting at a super-Eddington rate, which is also a known possible characteristic of accretion onto X-ray pulsars \citep{basko76}.

\subsection{Black holes}

\begin{figure*}
\includegraphics[width=1\textwidth]{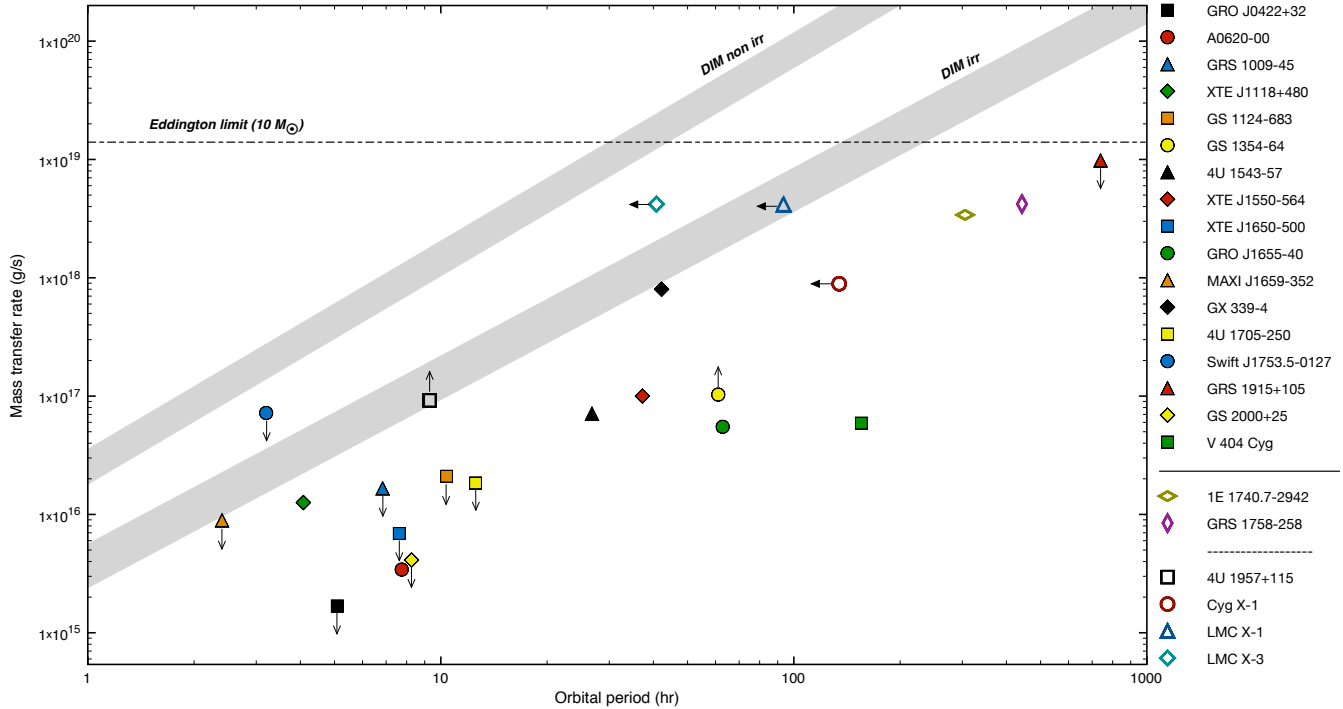} 
	\caption{Mass transfer rate as a function of the orbital period for XRBs with black holes. The transient and persistent sources have been indicated with filled and open 
	symbols respectively. The shaded grey areas indicated `DIM irr' and `DIM non irr' represent the separation between persistent (above) and transient systems (below) according to the disc instability model when, respectively, irradiation is taken into account and when it is neglected. The horizontal dashed line indicates the Eddington accretion rate for a 10 \msun\ black hole. All the upper limits on the mass transfer rate are due to lower limits on the recurrence time. The upper limits on the mass transfer rate of 4U 1957+115 and GS 1354-64 result from lower limits on the distance to the sources. The three left closed arrows do not indicate actual upper limits on the orbital period of Cyg X-1, LMC X-1 and LMC X-3. They emphasise that the radius of any accretion disk in these three high-mass XRBs is likely to be smaller than the one derived from the orbital period since they likely transfer mass by a (possibly focused) stellar wind instead of fully developed Roche lobe overflow. In the legend, the solid horizontal line separates transient and persistent systems. The dashed horizontal line stresses that the persistent nature of 1E 1740.7-2942 and GRS 1758-258 is unclear (see Section \ref{sec2}).}
	\label{dimbh}
\end{figure*}

The result of the test for the black hole sources is also in good agreement with the theoretical expectations. However, the low number of persistent sources, 
the uncertainties associated to their nature and the numerous upper or lower limits due to unconstrained recurrence time or distance, weaken the significance
of the test. Beyond this general assessment, we discuss below some individual cases that do not clearly fit into this global picture.

\subsubsection{1E 1740.7-2942 and GRS 1758-258}
 
 These two persistent sources should be transient, according to their position in the diagram.
 As already discussed in this paper, their classification as persistent sources is unclear and given their large orbital period, 
 decades-long outbursts are definitely not excluded. On the other hand, their orbital periods have been derived from modulations observed in their long-term 
 \rxte\ lightcurves \citep{smith02a} which might be less reliable than an optical determination. Consequently, we cannot exclude shorter orbital periods which
 could shift their position in the``persistent zone'' of the diagram.
For the above mentioned reasons, we won't consider these sources as contradictory to the DIM predictions. 
  
\subsubsection{The size of the disc in Cyg X-1}
 
We note on Fig.~\ref{dimbh} that Cyg X-1 lies below the stability line of the irradiated DIM. However, as we mentioned previously, the size of its accretion disc 
might be smaller than the one derived through its orbital period. If the persistent/transient division is correctly described by the irradiated DIM, we can estimate
an upper limit on the size of the disc of Cyg X-1, so that it is a persistent source given its mass transfer rate. Equating Eq. \ref{mcrit-irr} with our estimated \mext\
for Cyg X-1 and using a black hole mass of $14.8 \msun$ we obtain an outer disc radius of $3.6 \times 10^{11}$ cm. This implies that the accretion disc of Cyg X-1 should
be at least 3 times smaller than the size derived assuming Roche lobe overflow. 
 
\subsubsection{Is Swift J1753.5-0127 a new persistent source?} 

Swift J1753.5-0127 was discovered by the Burst Alert Telescope (BAT) on board Swift on 2005 May 30 \citep{palmer05} and, at the time of writing,
 is still undergoing the same outburst. No transition to the canonical soft state has been reported so far and the source always remained in hard states (Soleri et al. submitted). 
Given its particularly long outburst, its short orbital period (3.2h) and its position in the diagram of Fig. \ref{dimbh} (i.e. in the persistent zone)
 we investigated the possibility that Swift J1753.5-0127 is a ``new persistent system". 
 We estimated an upper limit to the mass of the disc at the onset of the outburst by assuming that the surface density at all radii is equal to the critical surface density
 required to trigger an outburst\footnote{This assumption lead to an upper limit on the mass of the disc since the only requirement to trigger an outburst according to the DIM is
 that the surface density is above the critical value at a single radius only.}. Given that the disc mass increases with radius, we assumed a 15 \msun\ black hole and
  obtained $M_{\rm disc} = 2.2 \times 10^{25} g$.
The 0.1-200 keV lightcurve integrated over time then gives us the total mass accreted from the start of the outburst until 2012 January 1: $M_{\rm accr} = 6.6 \times 10^{25} g$. 
To estimate $M_{\rm accr}$ we used the lower limit to the distance to the source (7.2 kpc; \citealt{zurita08}) and a radiative efficiency $\eta=0.1$. If the lower limit to the distance is
correct, $M_{\rm accr}$ should be considered as a lower limit to the total mass accreted since the fluence in the 0.1-200 keV range leads to a lower limit on the mass accreted. Consequently, even assuming that the entire (maximum) disc mass was consumed in the outburst, the lower limit of the mass accreted so far is 3 times higher and the outburst is still ongoing. If we assume that the difference between the mass accreted and the disc mass has been transferred by the donor star along the outburst, we derive a mass transfer rate of 
$\mext \sim 2 \times 10^{17} {\rm g \, s}^{-1}$ which is an order of magnitude above the critical mass accretion rate. Although very simple and crude, this calculation could indicate that
Swift J1753.5-0127 is a newly born persistent source. However, considering the extremely low probability of observing such an event, it is likely that the mass transfer rate
is in fact enhanced during the outburst by, e.g., irradiation of the donor star. If our calculation is realistic, the enhanced $\mext$ is greater than the critical accretion rate, which means that the outburst will last until the enhancement of the mass transfer weakens.

\subsection{A measure of the source activity - the transientness parameter}\label{tness}

From a theoretical point of view, for a given orbital period, we would expect transient systems located closer to the stability line to be more active, i.e. to undergo outbursts
more regularly. Indeed, for a given size of the disc, the time needed to ``refill'' the disc after an outburst and reach the critical density necessary to enter in a new outburst 
 will be lower if the mass transfer rate from the companion is higher. To test this idea we can define the ``transientness'' parameter of a source as the ratio between the estimated mass transfer rate and the critical accretion rate calculated for the orbital period of the source. Table \ref{bh-mdot} and \ref{ns-mdot} give the values of the 
 transientness for our selected sources.
 
 The transientness of different sources can then be compared with their average recurrence time. From the DIM point of view, we expect the average recurrence time
  to decrease with the transientness. However, the DIM, in its standard form, predicts a regular recurrence time and identical outbursts, which is not observed in the majority of cases. 
  In particular, some weak outbursts might be only echoes (i.e. XTE J1550-564, see below) of a ``real'' outburst as defined by the DIM.
 To estimate the average recurrence time of a given system, we therefore averaged the inter-outburst times weighted by the fluence of their subsequent outburst.  
 This should minimise the bias introduce by short recurrence time leading to weak outbursts.

Note that the black hole system XTE J1550-564 underwent a major outburst in 1998 (year of its discovery) followed by 4 weak outbursts, one every year, until 2003.
Since then, the source remained in quiescence. Estimating the average recurrence time is thus relatively complex in this case. In order to take into account the inter-outburst 
time preceding the major outburst in 1998, we calculated the time necessary to accumulate the mass that was accreted during the 1998 outburst using our estimate of the
mass transfer rate. We obtained $\sim 28$ years, which led to a weighted average recurrence time of 20 years when including the 4 others inter-outburst times.

\begin{figure}
\includegraphics[width=0.49\textwidth]{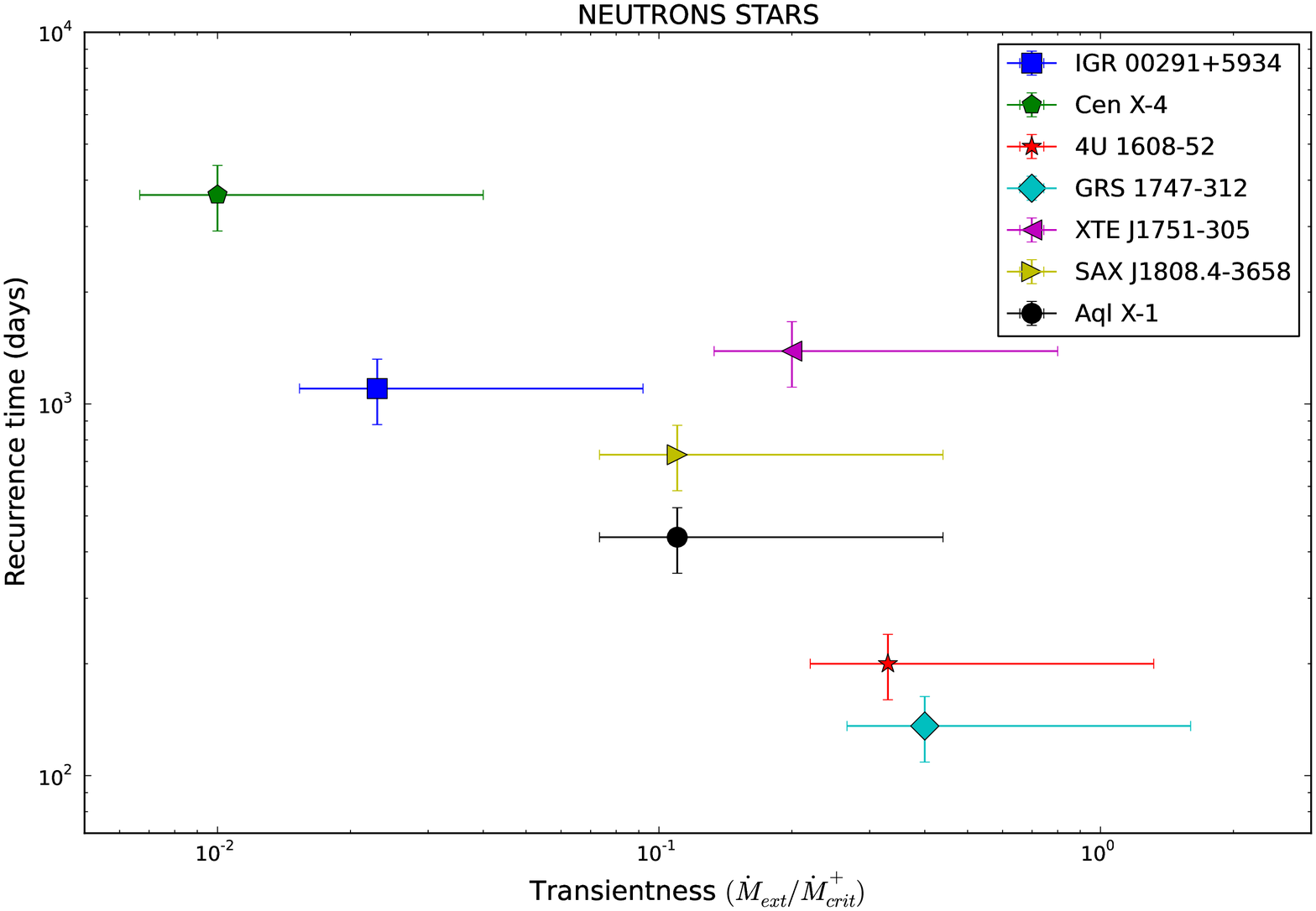} 
	\caption{Average recurrence time as a function of the transientness parameter ($\mext / \dot{M}_{\rm crit}$; see section \ref{tness}) for the neutron stars sample. }
	\label{dtdmns}
\end{figure}

\begin{figure}
	 \includegraphics[width=0.49\textwidth]{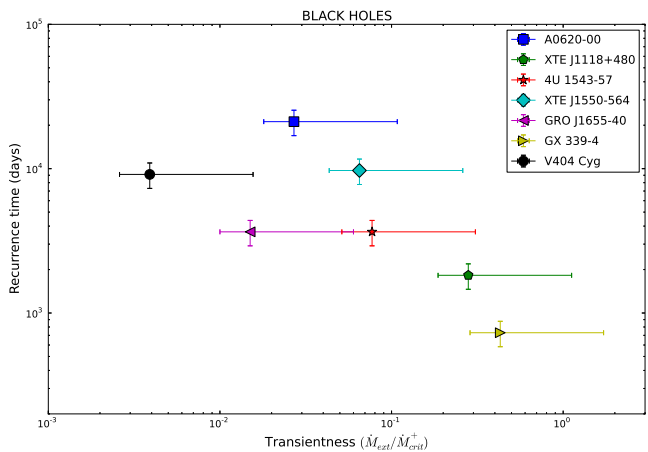} 
	\caption{Average recurrence time as a function of the transientness parameter ($\mext / \dot{M}_{\rm crit}$; see section \ref{tness}) for the  black holes sample. }
	\label{dtdmbh}
\end{figure}

 Fig. \ref{dtdmns} and \ref{dtdmbh} present the recurrence time as a function of the transientness for the neutron stars and the black holes respectively. 
 We have only selected sources for which the recurrence time is constrained by at least two outbursts. The error bars are presented for indicative purpose
 and correspond to uncertainties of a factor 3 on the mass transfer rate and of $20\%$ on the recurrence time. 
 Given the large uncertainties, we cannot draw firm and quantitative conclusions on the relation between the transientness and the recurrence time, but we note 
 clear trends in these two figures which are consistent with the DIM predictions: systems with higher transientness tend to undergo outbursts more frequently.
 This result strengthens again the DIM as the correct model to describe the global long-term activity of X-ray binaries.

\section{Summary and conclusion}

We have revisited and refined a simple yet fundamental test of the disc instability model for X-ray binaries, first carried out by \citet{van-paradijs96}.
Our work included a sample of 52 sources and confirmed that the DIM is able to explain the dichotomy between transient and persistent sources as long
as irradiation is taken into account. 

In the case of neutron stars, the data are in  precise agreement with the DIM predictions which allowed
us to constrain the irradiation constant between $10^{-2}$ and $10^{-3}$. Although less constrained by 
the data due to the low number of persistent sources and the numerous upper and lower limits, the black hole sources 
appear also in good agreement with the stability criterion predicted the DIM.

We then discussed some individual cases which do not clearly fit into this general conclusion, and mainly showed
that the classification of a source as a persistent system is arbitrary if the accretion disc is large enough to produce outburst longer than 
our historical record of its activity. We also emphasised the peculiar case of the black hole candidate Swift J1753.5-0127 which might have
already accreted more mass that the total maximum mass of its accretion disc. This points toward a mass transfer rate higher than the critical
accretion rate. 

Finally, we introduce the transientness parameter as a measure of the activity of a source and show a clear trend of the average recurrence time to decrease
with transientness in agreement with the DIM expectations. 

We therefore conclude that, despite some difficulties in reproducing the complex details of the light curves, the DIM succeeds in explaining the global behaviour 
of X-ray binaries averaged over a long enough period of time.

\section*{Acknowlegements}

MC would like to thank Dan Plant, Christian Knigge, Phil Charles, Tom Maccarone, Gabriele Ponti and Teo Mu\~noz-Darias for useful comments and discussions. 
GD acknowledges support from the European Community via contract ERC-StG-200911.

\bibliographystyle{mn2e}
\bibliography{/Users/Boss/TAFF/Biblio/biblio}

\bsp

\label{lastpage}

\end{document}